\documentclass{epsconf}
\usepackage{graphicx}
\usepackage{wrapfig}
\usepackage{amsmath}

\usepackage{siunitx}

\title{Mode Control of Electro-Magnetic Pulses in the VHF Band}
\author{\underline{M. Ehret}$^1$, L. Volpe$^1$, J.I. Apiñaniz$^1$, P. Puyuelo-Valdes$^1$, and G. Gatti$^1$}
\institute{$^1$ Centro de Laseres Pulsados (CLPU), 37185 Villamayor, Salamanca, Spain}

\begin{document}
\maketitle

We present experimental results for the controlled mitigation of electromagnetic pulses (EMP) produced in interactions of the \SI{1}{\peta\watt} high-power \SI{30}{\femto\second} Ti:Sa laser VEGA-3 with matter. This study aims at the band of very high frequencies (VHF), notably hundreds of \si{\mega\hertz}, comprising the fundamental cavity modes of the rectangular VEGA-3 vacuum chamber. We demonstrate mode suppression by tailoring of the laser-produced space charge distribution.

\section{Introduction}

Seed space charge fields are triggered by the interaction of the laser pulse that is focused to relativistic intensities onto a thin solid density aluminium (Al) target. As the VEGA laser pulse shows no pre-pulses capable of inducing a transparency of the target \cite{Volpe_2019}, the main acceleration mechanism of charged particles is Target Normal Sheath Acceleration (TNSA) \cite{Snavely_2000, Wilks_2001}. In TNSA, a population of laser-heated relativistic electrons escapes the target and the successive potential dynamics leads to the formation of sheath fields which are capable of accelerating ionized surface contaminants up to several tens of \si{\mega\electronvolt\per\atomicmassunit} \cite{Roth_2016}. Ions co-propagate with slow electrons and form a quasi-neutral beam which we will neglect in the following. With a focus on the \si{\mega\hertz} domain, the charging of the target on the \si{\pico\second}-timescale \cite{Poye_2018} is instantaneous: the target is initially at a net positive potential. Relativistic electrons propagate predominantly in laser-forward direction \cite{Rusby_2015} and yield an asymmetric charge separation. They distribute negative charge across the setup when they stop, on the \si{\nano\second}-timescale of their time of flight. Return currents rise on the same timescale of \si{\nano\second} \cite{Dubois_2014}, and resonances within the cavity can build up. The resulting broadband EMP bears the geometry of the experimental setup in its mode structure \cite{Consoli_2020, Nelissen_2020}.

\section{Experimental Study}

The laser pulse is compressed to \SI{30}{\femto\second} duration and focused via a $f/10$ off-axis parabola. The laser intensity reaches \SI{2.3e20}{\watt\per\square\centi\metre} on target, extrapolated from low-energy focus measurements. The normalized vector potential of the laser pulse calculates to $a_0 = 10.4$. The average energy $k_\text{B} T_\text{h}$ that is transferred to an electron of the relativistic population follows approximately \cite{Kluge_2011} with the ponderomotive scaling \cite{Wilks_2001} $k_\text{B} T_\text{h} = m_\text{e} c^2 ( \sqrt{1+a_0^2} - 1 ) = \SI{4.8}{\mega\electronvolt}$. Here $m_\text{e}$ is the electron rest mass and $c$ denotes the speed of light in vacuum.

\begin{figure}
\centering
\vspace{0cm} 
\begin{minipage}{.5\textwidth}
  \centering
  \includegraphics[width=70mm,trim={5mm 5mm 68mm 5mm},clip]{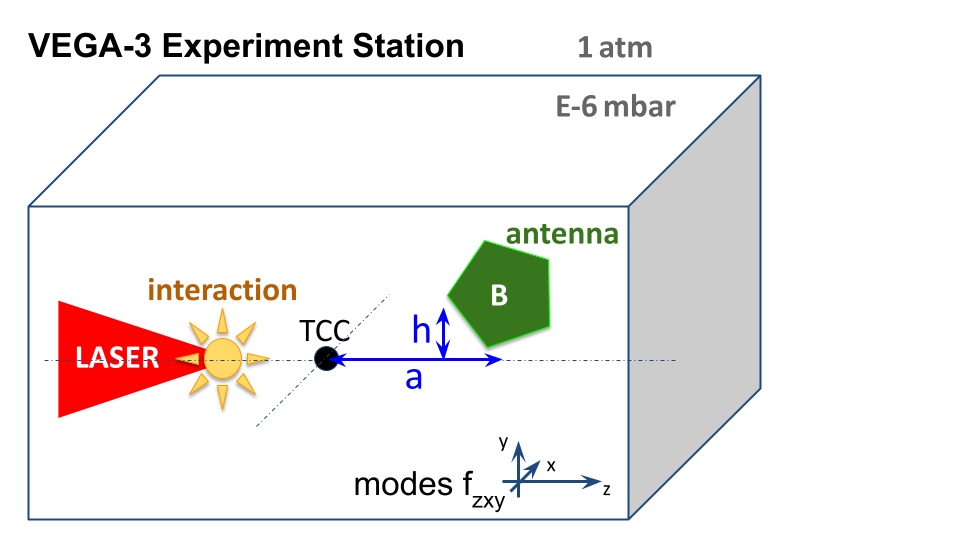}
  \caption{\it \small General Setup}
  \label{fig:VEGA3}
\end{minipage}%
\begin{minipage}{.5\textwidth}
  \centering
  \includegraphics[width=70mm,trim={5mm 5mm 68mm 5mm}]{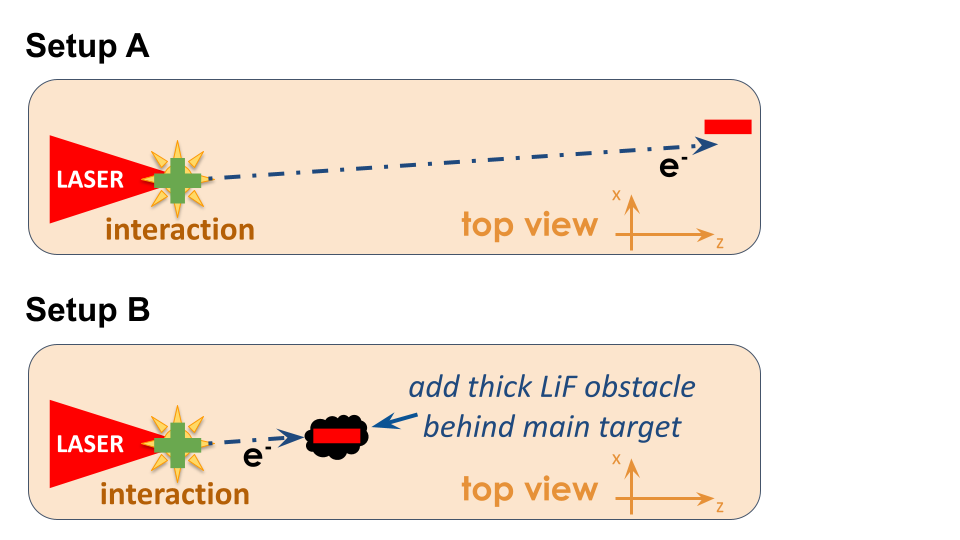}
  \caption{\it \small Studies}
  \label{fig:studyEMP}
\end{minipage}%
\vspace{0cm} 
\end{figure}

The \SI{3}{\micro\metre} thick solid density Al target is placed in the laser focus position and vertically inclined by \SI{12.5}{\degree} with respect to the laser axis. The interaction point is \SI{31}{\centi\metre} off-centre with respect to the rectangular vacuum chamber, towards the parabola. The setup is shown in figure (Fig.~\ref{fig:VEGA3}). The inclination of the target leads to a deposition of negative charge in one corner of the interaction chamber, see figure (Fig.~\ref{fig:studyEMP} A). The relativistic electron beam is completely stopped in the \si{\centi\metre} thick wall of the vacuum chamber. Transverse and longitudinal electric fields are likely to build up. Return currents in the skin depth of the chamber walls then stream longitudinally and radially, causing a transverse azimuthal and radially toroidal magnetic field. The excited modes are expected to be $f_{100}$, $f_{001}$, and higher orders. The positive potential of the target and radial return currents can excite modes such as $f_{101}$, $f_{010}$, $f_{110}$, $f_{111}$, as well as higher orders.

In a medium with speed of light $c$, resonant modes have frequencies
\begin{equation}
\label{eq:frequencies}
    f_\mathrm{mnp} = \frac{c}{2} \cdot \sqrt{\left(\frac{m}{\Delta x}\right)^2+\left(\frac{n}{\Delta y}\right)^2+\left(\frac{p}{\Delta z}\right)^2}\quad ,
\end{equation}
where $\{m,n,p\}$ are the mode numbers of the corresponding dimensions $\{\Delta x,\Delta y,\Delta z\}$. The width in laser-transverse direction is $\Delta x \in [ \SI{120}{\centi\metre} ; \SI{169}{\centi\metre} ] $, the height is $\Delta y \in [ \SI{60}{\centi\metre} ; \SI{70}{\centi\metre} ] $, and the length in laser-forward direction is $\Delta z \in [ \SI{164}{\centi\metre} ; \SI{175}{\centi\metre} ] $. Here the lower bound is given by the rectangular inner surface. The outer bound is given by ports, flanges and window surfaces which are not expected to play a major role for resonances but will contribute to losses.

The detection of EMP is achieved by a passive calibrated B-field antenna with large bandwidth from \SI{9}{\kilo\hertz} to \SI{400}{\mega\hertz} (Aaronia MDF9400). The magnetic field antenna is positioned inside the cavity, in the vertical plane of the laser propagation, longitudinally $a=\SI{17.5}{\centi\metre}$ behind the centre of the cavity and $h=\SI{17.5}{\centi\metre}$ above the horizontal centre-plane. The signal is transported via calibrated double shielded SMA cables and through a floating feed-through. The conductive connection of SMA cables and ground is avoided completely. Waveforms are captured with an oscilloscope of \SI{1}{\giga\hertz} bandwidth and $2.5\,\text{GS/s}$ sampling rate.

The relevant cavity modes are $f_{001} = \SI{91}{\mega\hertz}$, $f_{100} = \SI{125}{\mega\hertz}$, $f_{101} = \SI{155}{\mega\hertz}$, $f_{002} = \SI{183}{\mega\hertz}$, $f_{010} = \SI{250}{\mega\hertz}$, $f_{200} = \SI{250}{\mega\hertz}$, $f_{011} = \SI{266}{\mega\hertz}$, $f_{003} = \SI{274}{\mega\hertz}$, $f_{110} = \SI{279}{\mega\hertz}$, $f_{111} = \SI{294}{\mega\hertz}$, $f_{202} = \SI{310}{\mega\hertz}$, $f_{112} = \SI{334}{\mega\hertz}$, $f_{211} = \SI{365}{\mega\hertz}$, $f_{004} = \SI{366}{\mega\hertz}$, $f_{203} = \SI{371}{\mega\hertz}$, $f_{300} = \SI{375}{\mega\hertz}$, and $f_{212} = \SI{398}{\mega\hertz}$.

\begin{figure}
\centering
\vspace{0cm} 
\begin{minipage}{\textwidth}
  \centering
  \includegraphics[width=140mm,trim={5mm 92mm 68mm 5mm},clip]{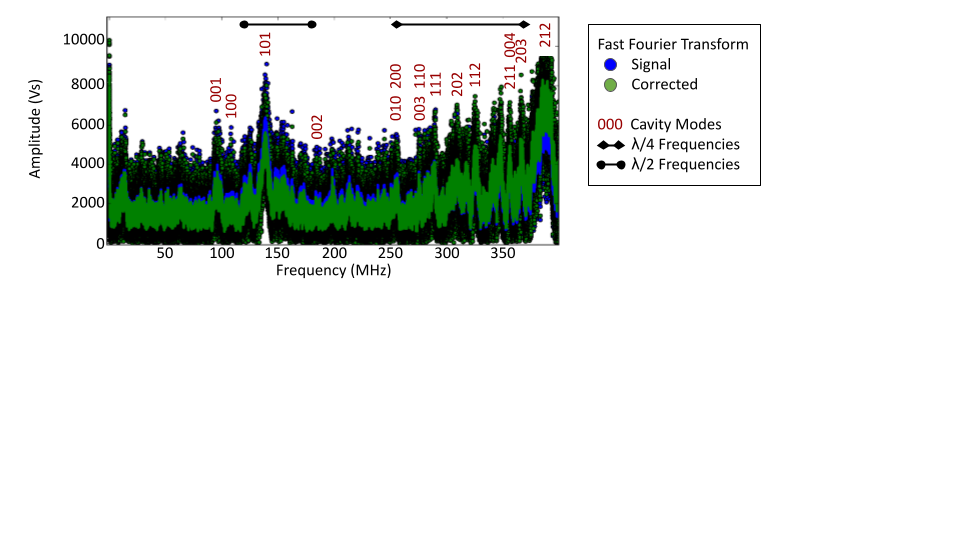}
  \caption{\it \small Time integrated spectrum of the magnetic field measured in the VEGA-3 interaction chamber for shots on solid targets, with indication of identified rectangular cavity modes and the resonance of the vertical stalks that hold optics ($\lambda/4$).}
  \label{fig:iniEMP}
\end{minipage}%
\vspace{0cm} 
\end{figure}

The Fast Fourier Transform (FFT) of the resulting waveform data acquired during the experiment is shown in figure (Fig.~\ref{fig:iniEMP}). One notes the strong mode $f_{101}$ selected by the initial diagonal space charge fields. The respective fundamentals $f_{100}$ and $f_{001}$ are clearly visible. In addition to the cavity modes, cylindrical metallic $1/2$-inch mounts of optics with length ranging from \SIrange{20}{30}{\centi\metre} produce spectral peaks with antenna fields in a range from \SIrange{250}{372}{\mega\hertz}. Their resonance frequencies are defined by a match of multiples of $\lambda/4$ with the length of the grounded monopole rod. Above this frequencies, there is a clear peak at \SI{385 \pm 10}{\mega\hertz}. This corresponds to \SI{19 \pm 1}{\centi\metre} and might be related to the motor block holding the target.

In an attempt to alter the mode structure, a beam dumper from LiF of \SI{\approx 1.5}{\centi\metre} thickness and \SI{\approx 2.5}{\centi\metre} diameter is positioned at \SI{1}{\centi\metre} behind the Al target, see figure (Fig.~\ref{fig:studyEMP} B). The crystal acts as electron catcher, removing part of the forward propagating space charge and thus weakening the fundamental modes in transverse and longitudinal direction. The stopping of electrons was simulated with the Monte Carlo simulation code CASINO \cite{Drouin_2007} for a LiF crystal of density \SI{1.06756}{\gram\per\cubic\centi\metre}. It stops all electrons up to an energy of \SI{2.8}{\mega\electronvolt}.

\begin{figure}
\centering
\vspace{0cm} 
\begin{minipage}{\textwidth}
  \centering
  \includegraphics[width=140mm,trim={5mm 92mm 68mm 5mm},clip]{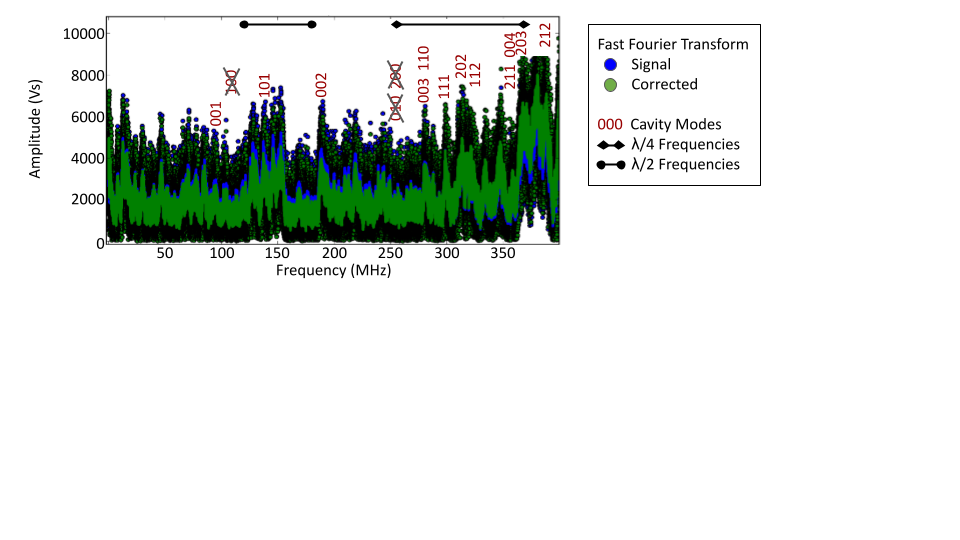}
  \caption{\it \small Time integrated spectrum of the magnetic field measured in the VEGA-3 interaction chamber for shots on solid targets with inserted LiF beam dump shortly behind the solid target, with indication of identified rectangular cavity modes and the resonance of the vertical stalks that hold optics ($\lambda/4$).}
  \label{fig:finEMP}
\end{minipage}%
\vspace{0cm} 
\end{figure}

Results shown in figure (Fig.~\ref{fig:finEMP}) show a reduction of the fundamental mode in longitudinal direction, a removal of the fundamental mode in transverse direction and a clear reduction of energy in mode $f_{101}$. The return current in the beam dumper mount of about \SI{20.5}{\centi\metre} height manifests in a spectral peak that appears at \SI{366 \pm 4}{\mega\hertz} -- on the left edge of the large peak that might be attributed to the motor block. The modification of the experimental geometry allows to influence the mode structure in the cavity, by implicitly selecting the favoured modes which will become excited.

\section{Summary}

The excited cavity modes are clearly detected by the compact antenna system. This work shows that the mode structure can be tailored by a modification of the experimental geometry. VHF EMP are relevant in science and technology in many fields ranging from homeland security \cite{Baker_2019} to medicine \cite{Kim_2019} for their potential hazards and promising applications. Their controlled generation will pave the ground for studies in those fields.

\section{Acknowledgements}

This work was supported by the European IMPULSE project with funding from the European Union’s Horizon 2020 research and innovation program under grant agreement No 871161.

\end{document}